\documentclass[reprint,superscriptaddress,hyphens,amsmath,amssymb,aps,prx,float]{revtex4-2}
\usepackage[T1]{fontenc}
\usepackage[latin9]{inputenc}
\usepackage{graphicx}% Include figure files
\usepackage{dcolumn}% Align table columns on decimal point
\usepackage{bm}% bold math
\usepackage{amsmath}
\usepackage{physics}
\usepackage[usenames,dvipsnames]{color}
\usepackage{xurl}
\usepackage[breaklinks]{hyperref}
\usepackage{hypernat}
\usepackage{breakurl}
\PassOptionsToPackage{hyphens}{url}
\hypersetup{colorlinks, citecolor=Blue, linkcolor=Blue, urlcolor=Blue}

\begin{document}
\title{High-precision mapping of diamond crystal strain using quantum interferometry}
\author{Mason C. Marshall}
\email[Current address: Time and Frequency Division, NIST, Boulder, Colorado, 80305, USA.  ]{mason.marshall@nist.gov}
\affiliation{Department of Electrical and Computer Engineering, University of Maryland, College Park, Maryland, 20742, USA}
\affiliation{Center for Astrophysics $\vert$ Harvard \& Smithsonian, Cambridge, Massachusetts, 02138, USA}
\affiliation{Quantum Technology Center, University of Maryland, College Park, Maryland, 20742, USA}
\author{Reza Ebadi}
\affiliation{Quantum Technology Center, University of Maryland, College Park, Maryland, 20742, USA}
\affiliation{Department of Physics, University of Maryland, College Park, Maryland, 20742, USA}
\author{Connor Hart}
\affiliation{Quantum Technology Center, University of Maryland, College Park, Maryland, 20742, USA}
\author{Matthew J. Turner}
\affiliation{Quantum Technology Center, University of Maryland, College Park, Maryland, 20742, USA}
\author{Mark J.H. Ku}
\affiliation{Quantum Technology Center, University of Maryland, College Park, Maryland, 20742, USA}
\affiliation{Department of Physics and Astronomy, University of Delaware, Newark, Delaware, 19716, USA}
\affiliation{Department of Materials Science and Engineering, University of Delaware, Newark, Delaware, 19716, USA}
\author{David F. Phillips}
\affiliation{Center for Astrophysics $\vert$ Harvard \& Smithsonian, Cambridge, Massachusetts, 02138, USA}
\author{Ronald L. Walsworth}
\email{walsworth@umd.edu}
\affiliation{Department of Electrical and Computer Engineering, University of Maryland, College Park, Maryland, 20742, USA}
\affiliation{Center for Astrophysics $\vert$ Harvard \& Smithsonian, Cambridge, Massachusetts, 02138, USA}
\affiliation{Quantum Technology Center, University of Maryland, College Park, Maryland, 20742, USA}
\affiliation{Department of Physics, University of Maryland, College Park, Maryland, 20742, USA}

\date{Draft: \today}

\begin{abstract}
Crystal strain variation imposes significant limitations on many quantum sensing and information applications for solid-state defect qubits in diamond.  Thus, precision measurement and control of diamond crystal strain is a key challenge.  Here, we report diamond strain measurements with a unique set of capabilities, including micron-scale spatial resolution, millimeter-scale field-of-view, and a two order-of-magnitude improvement in volume-normalized sensitivity over previous work \cite{StrainPaper}, reaching $5(2) \times 10^{-8}/\sqrt{\rm{Hz}\cdot\rm{\mu m}^{-3}}$ (with spin-strain coupling coefficients representing the dominant systematic uncertainty).  We use strain-sensitive spin-state interferometry on ensembles of nitrogen vacancy (NV) color centers in single-crystal CVD bulk diamond with low strain gradients.  This quantum interferometry technique provides insensitivity to magnetic-field inhomogeneity from the electronic and nuclear spin bath, thereby enabling long NV ensemble electronic spin dephasing times and enhanced strain sensitivity, as well as broadening the technique's potential applications beyond isotopically-enriched or high-purity diamond.  We demonstrate the strain-sensitive measurement protocol first on a scanning confocal laser microscope, providing quantitative measurement of sensitivity as well as three-dimensional strain mapping; and second on a wide-field imaging quantum diamond microscope (QDM).  Our strain microscopy technique enables fast, sensitive characterization for diamond material engineering and nanofabrication; as well as diamond-based sensing of strains applied externally, as in diamond anvil cells or embedded diamond stress sensors, or internally, as by crystal damage due to particle-induced nuclear recoils.
\end{abstract}

\maketitle
\newcommand{\w}{3.5in}

\newcommand{\introfigure}{
\begin{figure}[htbp!]
\begin{center}
\includegraphics*[width=\columnwidth]{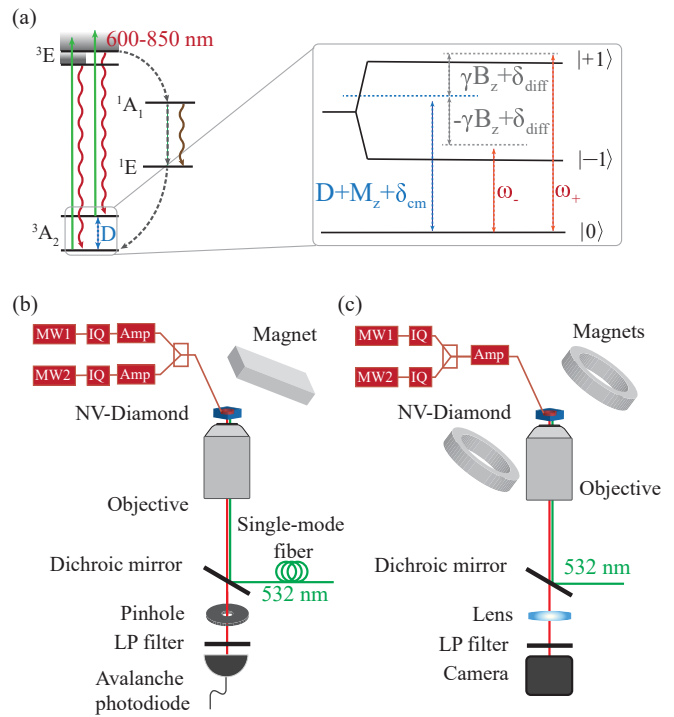}
%introfigure_v4.eps}
\caption{\textbf{(a)} Simplified NV center level structure.  
%The electronic spin is initialized and read out via application of green light and collection of spin-dependent red fluorescence.  
Detail shows the ground-state electronic spin energy levels, as well as microwave (MW) field frequencies and detunings used in strain spectroscopy.  (Additional details in text.)   \textbf{(b)} and \textbf{(c)} Schematics of the confocal microscope and widefield quantum diamond microscope (QDM) used in this work - see Appendices \ref{sec:confocalapparatus} and \ref{sec:QDMapparatus} for additional details.}
\label{fig:introfigure}
\end{center}
\end{figure}
}

\newcommand{\sequencefig}{
\begin{figure}[htbp!]
\begin{center}
\includegraphics*[width=\columnwidth]{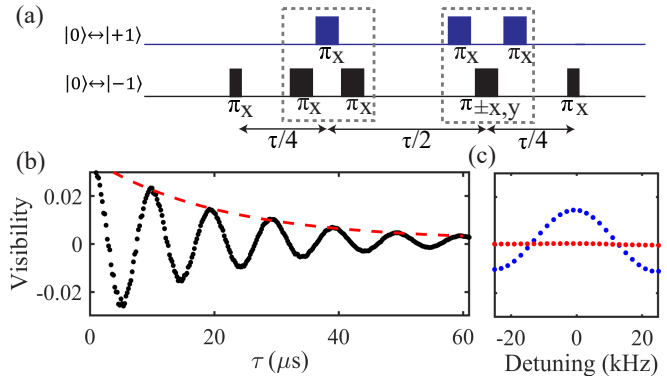}
\caption{\textbf{(a)} Timing diagram of microwave (MW) pulses applied during a strain-CPMG measurement. \textbf{(b)} strain-CPMG measurement of NV ensemble visibility $\nu$ (defined in equation \eqref{eqn:suppvis}) as a function of evolution time, for MW detuning $\delta_{\rm{cm}}\approx$ 0.1 MHz (see text for details).  Red dashed line gives the decay envelope, with amplitude $A e^{-\tau/T_D}$ (see text).  The strain-CPMG dephasing time $T_D$ is 21 ${\rm \mu s}$, compared to the canonical inhomgeneous dephasing time $T_2^\star = 7.5 ~{\rm \mu s}$ measured using a Ramsey sequence (see Appendix \ref{sec:dephasingtimes}).  \textbf{(c)} Calibration curves for fixed evolution time $\tau_1 = 21~{\rm \mu s}$, showing $\nu$ as a function of applied detunings $\delta_{\rm cm}$ (blue, calibrating strain- or temperature-induced shifts in $D$) and $\delta_{\rm diff}$ (red, calibrating magnetic field variation).}
\label{fig:sequencefig}
\end{center}
\end{figure}
}

\newcommand{\confocalfig}{
\begin{figure}[htbp!]
\begin{center}
\includegraphics*[width=\columnwidth]{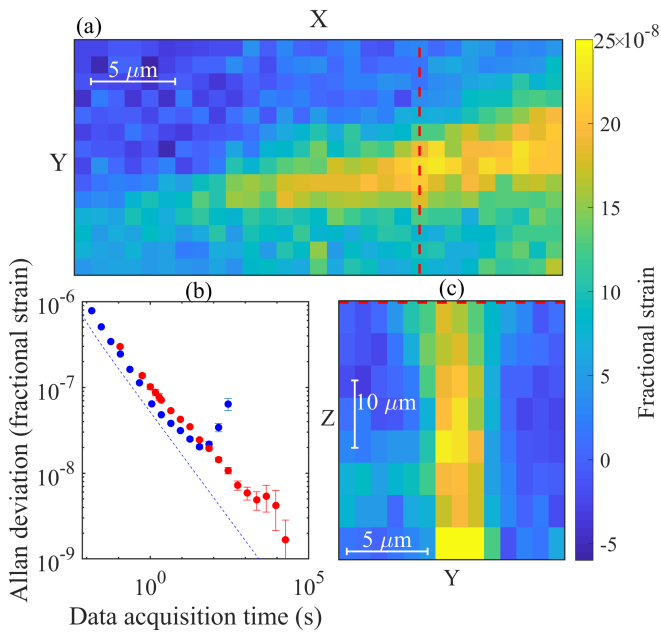}
\caption{Three-dimensional NV-diamond strain measurements with a scanning confocal microscope. \textbf{(a)} Strain structure measured at the diamond surface. Position of slice in panel (c) is marked with a red dashed line. \textbf{(b)} Allan deviation for confocal strain measurements.   Blue points are for single-position measurements, while red are for the ``gradiometry'' configuration (see text).  Blue dotted line represents the calculated limit from shot noise in the avalanche photodiode (APD) used to measure NV fluorescence (see Appendix \ref{sec:noiselimit} for details). \textbf{(c)} Orthogonal slice across the strain feature, scanning beneath the diamond surface (marked by a red dashed line).  Spatial scales are unequal due to the high index of refraction of diamond.}
\label{fig:confocalfig}
\end{center}
\end{figure}
}

\newcommand{\imageronefig}{
\begin{figure*}[htbp!]
\begin{center}
\includegraphics*[width=\textwidth]{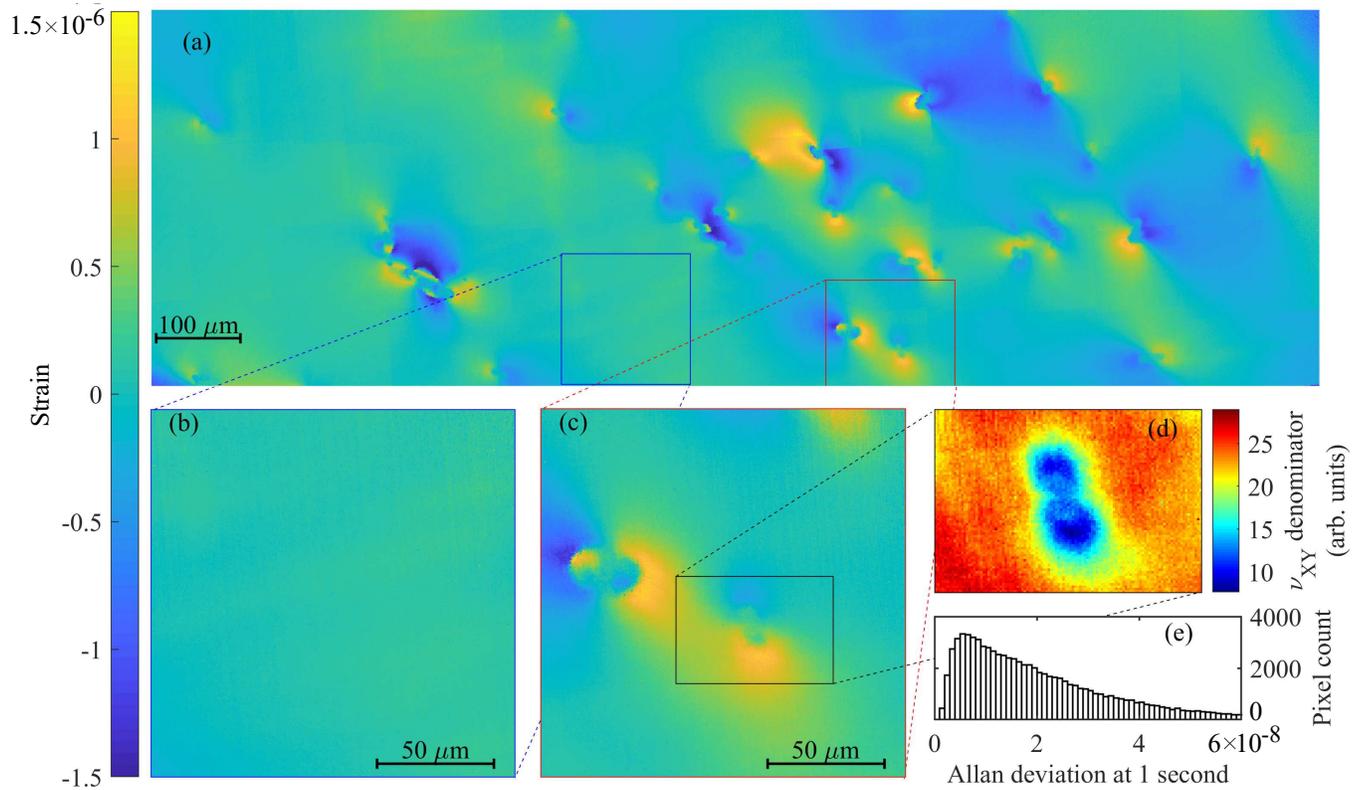}
\caption{Widefield strain images taken with a quantum diamond microscope (QDM).  \textbf{(a)} Wide survey of strain features on diamond section B.  Individual fields-of-view, each representing about 150 $\times$ 150 $\mu$m$^2$ and one second of data acquisition, were manually stitched together by aligning overlapping features to form a composite image; minor compositing artifacts result from temperature drift between image acquisitions.  \textbf{(b)} and \textbf{(c)} Single fields-of-view, as measured in one second of data acquisition each. Panel \textbf{(b)} features extremely homogeneous strain, characteristic of this diamond material. Panel \textbf{(c)} showcases two strain features, likely induced by crystallographic defects incorporated during CVD growth.  \textbf{(d)} The denominator of the XY-normalized visibility indicates areas where intra-pixel strain gradients degrade the dephasing time $T_D$ (see text).  \textbf{(e)} Histogram of Allan deviations for 1 second of data acquisition for each pixel in panel (b), giving an estimate of the strain measurement uncertainty.}
\label{fig:imageronefig}
\end{center}
\end{figure*}
}

\newcommand{\smfive}{
\begin{figure*}[hbtp!]
\begin{center}
\includegraphics*[width=\textwidth]{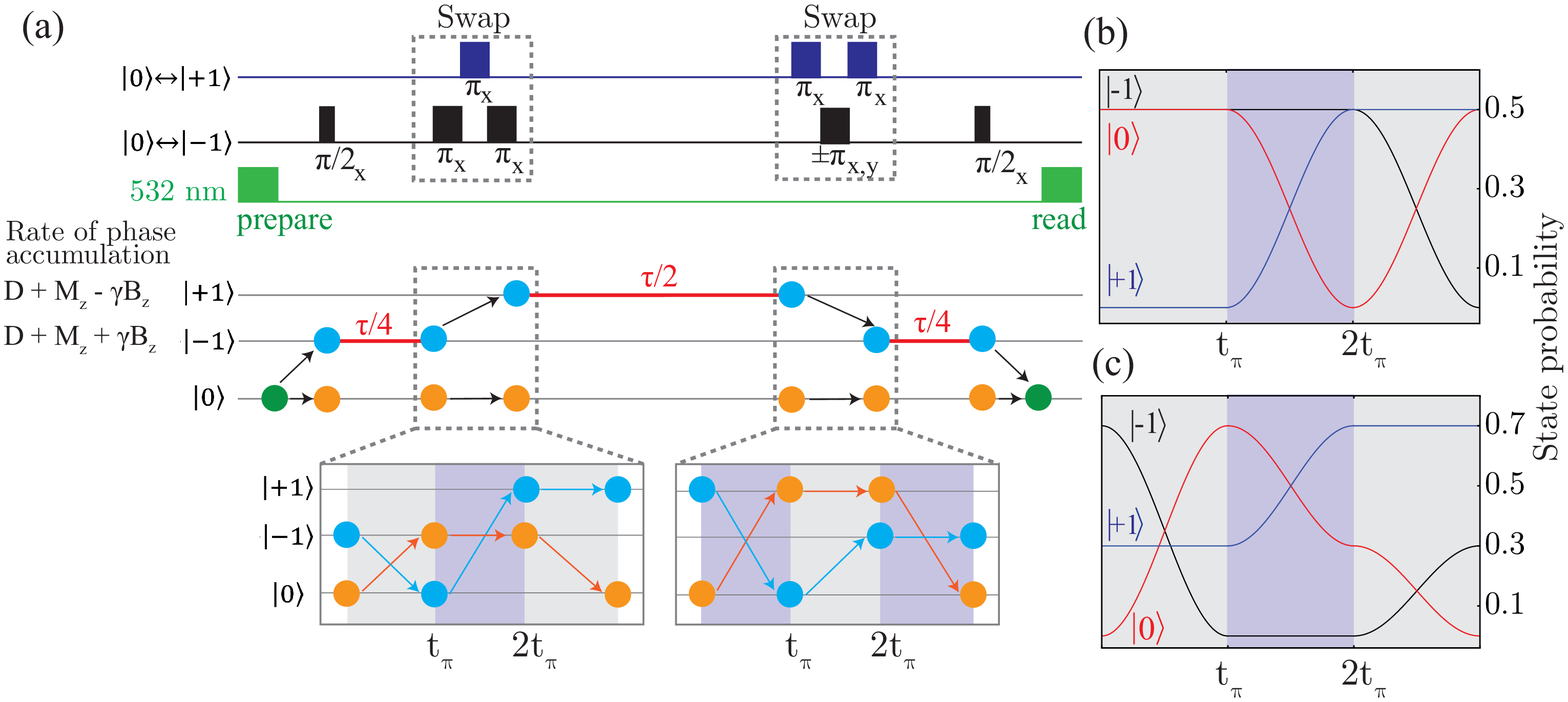}
\caption{\textbf{(a)} Detailed schematics of the strain-CPMG interferometric sensing protocol. Top: strain-CPMG pulse sequence. Bottom: NV spin population in each of the $\ket{{\rm m_s = 0,\pm 1}}$ states as a function of time. The swap operation exchanges the population between $\ket{+1}$ and $\ket{-1}$ states. Free evolution intervals are highlighted in red. Insets schematically illustrate evolution of the NV spin-state populations during swap operations. The first laser pulse initializes the NVs into $\ket{0}$, while the final laser pulse reads out the state-dependent fluorescence. \textbf{(b,c)} Time evolution of the state populations under the swap operation, using Eq. \eqref{eqn:swap_evolution}. Gray and blue shaded areas show evolution under resonant MW fields addressing the $\ket{0} \leftrightarrow \ket{-1}$ and $\ket{0} \leftrightarrow \ket{+1}$ transitions, respectively.  \textbf{(b)} Analytical calculation of the time-course of NV spin-state populations undergoing the strain-CPMG protocol, for an initial condition chosen such that the NV spin-state populations are equally separated between $\ket{0}$ and $\ket{-1}$, similar to an ideal case for the first swap pulse in the experiment. \textbf{(c)} Analytical calculation for an arbitrary initial condition with  70\% of the NV population in $\ket{-1}$ and 30\% in $\ket{+1}$, such that the non-equal populations at each step better illustrates the dynamics under the swap pulses.}
\label{fig:SMfig1}
\end{center}
\end{figure*}
}

\newcommand{\smsix}{
\begin{figure}[htbp!]
\begin{center}
\includegraphics*[width=\columnwidth]{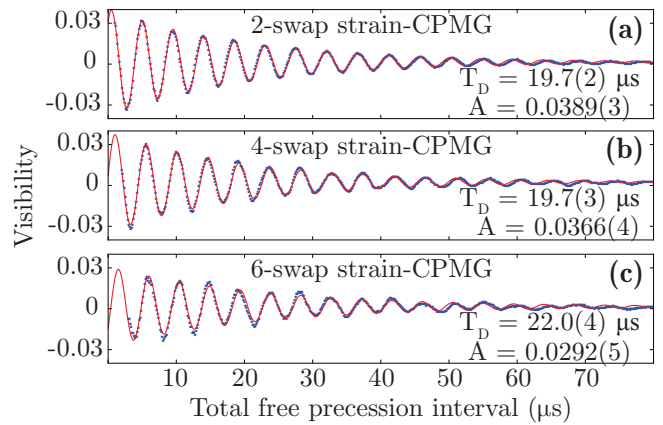}
\caption{Comparison of strain-CPMG sequences with different numbers of swap operations: time traces of the visibility $\nu$ (see equation \eqref{eqn:suppvis} below), with both MW tones detuned from resonance with a common-mode detuning $\delta_{\rm cm} = 0.35~{\rm MHz}$, for {\bf (a)} 2-swap, {\bf (b)} 4-swap, and {\bf (c)} 6-swap strain-CPMG sequences. Blue data points are measured values, while the red curves are fits to an exponentially decaying sinusoid. The two resulting fit parameters that determine sensitivity, i.e., the amplitude $A$ and decay rate $T_D$, are quoted for each sequence. The initial lag for the first data point arises from the minimum initial delay between swap operations.}
\label{fig:SMfig2}
\end{center}
\end{figure}
}

\newcommand{\smseven}{
\begin{figure}[htbp!]
\begin{center}
\includegraphics*[width=\columnwidth]{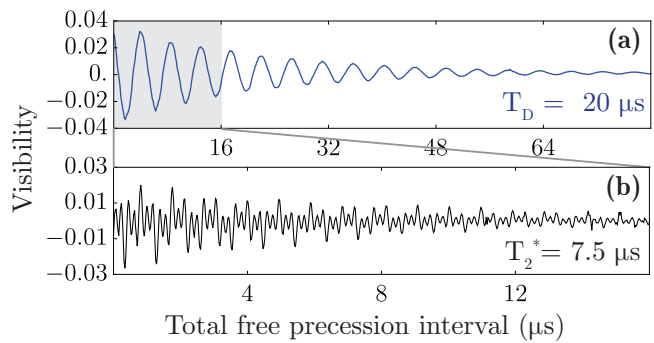}
\caption{Comparison of free induction decays for different measurement protocols. {\bf (a)} Visibility ($\nu$) as a function of evolution time for the strain-CPMG sequence with common MW detuning $\delta_{\rm cm} = 0.35 ~{\rm MHz}$, yielding a determination of $T_D$. {\bf (b)} SQ measurement, addressing the $\ket{0} \leftrightarrow \ket{+1}$ transition with 3 MHz detuning, providing a determination of $T_2^\star$. The visibility oscillates as a superposition of three frequencies, each of which are detunings from hyperfine splittings.  The SQ measurement result is shown for the first 16 ${\rm \mu s}$, addressing the same period as that shown in the gray shaded region in (a) for the strain-CPMG measurement result. %{\bf (c)} DQ measurement with 3 MHz detuning from each transition frequency. $T_2^*\{\rm DQ\}$ is the shortest dephasing time, indicating a large magnetic field inhomogeneity throughout the interrogated NV ensemble.
}
\label{fig:SMfig3}
\end{center}
\end{figure}
}

\newcommand{\smeight}{
\begin{figure*}[hbtp!]
\begin{center}
\includegraphics*[width=0.9\textwidth]{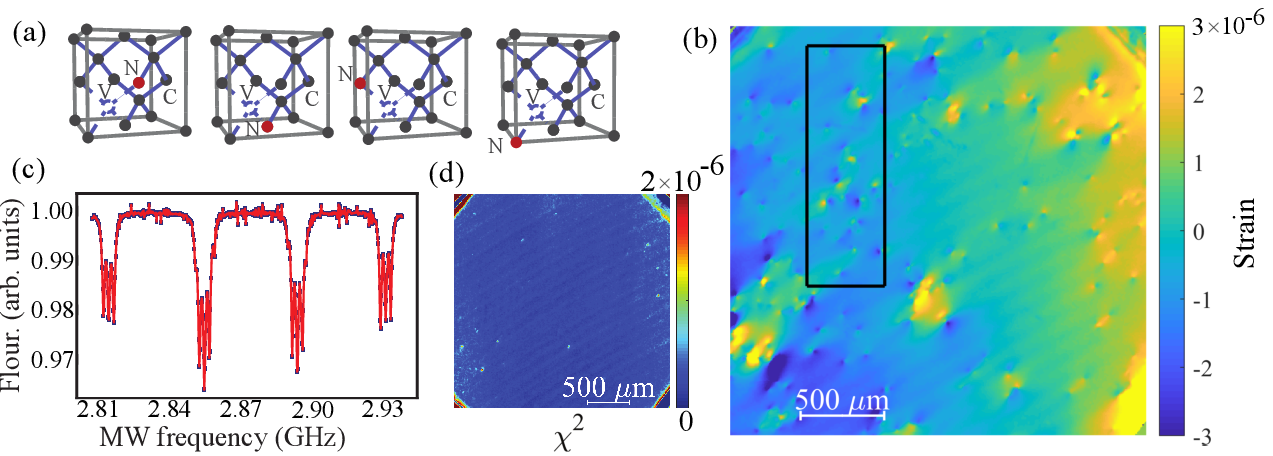}
\caption{\textbf{(a)} Four NV orientation classes, pointing along four crystallographic directions in the diamond crystal lattice.  \textbf{(b)} Widefield strain map of diamond section B, acquired using methods and QDM apparatus described in \cite{StrainPaper}.  The black rectangle designates the region shown in Fig.~\ref{fig:imageronefig}.  {\bf (c)} Example optically detected magnetic resonance (ODMR) spectrum from an NV ensemble; changes in fluorescence intensity reveal spin transition frequencies. Each NV resonance, corresponding to one or more of the four NV orientation classes, is split into three lines due to hyperfine interactions with the spin-1 $^{14}$N nucleus.  \textbf{(d)} Map of $\chi^2$ goodness-of-fit statistic for per-pixel fits yielding the strain map of panel (b).  Large intra-pixel strain gradients degrade the fits via deviation from Lorentzian lineshapes.}
\label{fig:SMfig_odmr}
\end{center}
\end{figure*}
}

\section{Introduction}

Quantum  defects in diamond are a leading platform for many applications, ranging from magnetometry and magnetic imaging \cite{QDMreview,SingleNVMagSensing2008,NVimaging2008,NVMagnetometer2008,RondinMagnetometry2014,NVNanoReview2014,SensitivityReport,ArunkumarSABRE2021} to quantum networking and communication \cite{PompiliQuantumNetwork2021,BhaskarQuantumCommunication2020}.  Such applications leverage the long spin coherence times and straightforward optical control of diamond defect qubits \cite{BalasubramanianCoherenceTime2009,DohertyPhysicsReports2013}, as well as the capability to engineer diamond with tunable defect density and geometry \cite{SensitivityReport,PurpleDiamond}, and to integrate qubits into photonic nanostructures \cite{NanophotonicPlatform2016,NguyenLukinNanophotonicRegister2019}.  These applications also critically depend on the host environment; e.g., diamond crystal strain inhomogeneity limits magnetometry by broadening and shifting qubit spin transitions \cite{StrainPaper}, degrades the decoherence of ensemble and single qubits \cite{SensitivityReport,vanDamEnglundImplantedNVCoherence2019,SiVStrain2018}, and spectrally broadens or quenches qubits near fabricated nanostructures \cite{KnauerFabStrain2020}.  Sensitive measurement of diamond strain is thus a key enabling technique for quantum applications, and as such has been the subject of significant recent study \cite{StrainPaper, AusStrain2019, EnglundStrain2016}.

Beyond this negative role, diamond strain can also have utility in quantum applications.  Control of qubit couplings via the strain environment can enable quantum metrology or networking \cite{SiVStrain2018,BennettLukinPhononSqueezing2013,XuQuantumInfoStrain2019,GolterOptomechanic2016} or drive qubit transitions \cite{MacQuarrieMechanicalControl2013,ChenAcousticDriving2020}.   Strain-mediated couplings also underlie hybrid quantum-nanomechanical systems \cite{BarsonNanomechanicalSensing2017,BarfussHybridSystems2019}.  Small diamonds as in situ strain sensors can be widely deployed as in high-pressure diamond anvil cells or in curing amorphous solids \cite{HoAnvilCells2020,HoAmorphousSolids2021}.  Finally, sensitive measurements of damage-induced strain in diamond are essential to proposed diamond-based particle detectors, such as for directional detection of dark matter consisting of weakly interacting massive particles (WIMPs) \cite{proposal,MarshallQST2021}.

Improvements in sensitive diamond strain measurement are therefore of broad importance.  Several techniques have been developed for the measurement of diamond crystal strain; however, quantum applications require fast sensing of small strains at relevant length scales, demanding state-of-the-art volume-normalized sensitivity.  X-ray tomography and microscopy require long data acquisition times (and, frequently, synchrotron beamline access) \cite{Friel2009,MooreXray2009,ourXray};  Raman spectroscopy faces a detection noise floor near strain of $10^{-5}$, far from the state of the art \cite{RamanStrain2011,AlexandermicroRaman1994}; and birefringence imaging integrates light through an entire diamond, limiting spatial resolution \cite{HoaBirefringence2014,StrainPaper}.

Here, we report strain measurements with record volume-normalized sensitivity of $5(2)\times10^{-8}/\sqrt{{\rm Hz \cdot \mu m}^{-3}}$ (with spin-strain coupling coefficients representing the dominant systematic uncertainty) -- an improvement by two orders of magnitude over previous work \cite{StrainPaper}.  Using spin-state interferometry of an ensemble of nitrogen vacancy (NV) centers, we achieve high precision via a strain-sensitive measurement protocol designed to be insensitive to static and low-frequency magnetic fields.  We apply this procedure to single-crystal CVD bulk diamond material grown by Element Six Ltd. with 99.995\% $^{12}$C and an estimated NV density of approximately 0.4 parts per million (ppm), featuring low strain and a long ensemble NV spin dephasing time $T_2^\star=7.5 \mu s$ (see Appendix \ref{sec:dephasingtimes} for additional details).  We first study strain in this diamond using a confocal scanning laser microscope; the known confocal volume allows us to quantitatively characterize the volume-normalized strain sensitivity and to demonstrate three-dimensional mapping of strain features in the diamond with micron-scale resolution.  We then apply the interferometry protocol using a wide-field quantum diamond microscope (QDM) \cite{GlennGGG,GlennMTB2015,QDMreview}, which provides high-sensitivity strain imaging of a millimeter-scale diamond section.  These results provide a benchmark for sensitive mapping of diamond strain with a unique combination of spatial resolution, field-of-view, and sensitivity; they thereby open a path to applications such as characterization and engineering of diamonds for quantum information and metrology, in situ strain measurements using micron-scale diamonds in soft materials, and a future dark matter search \cite{proposal,MarshallQST2021}.

We note that, while the record strain sensitivity achieved here leverages a high density of NV centers, this technique is generally applicable to diamonds with any NV density.  All else being equal, the sensitivity improves with the square root of the NV density.  Additionally, while these measurements used isotopically purified diamond, the insensitivity of the quantum interferometry technique to magnetic field inhomogeneity will enable precise strain measurements in diamonds with higher impurity or $^{13}$C content -- expanding the utility of strain sensing beyond specially engineered quantum diamonds.

\introfigure

\section{NV center spectroscopy overview}

Traditional techniques for measurement of diamond strain are limited in volume-normalized sensitivity by a higher noise floor, coarser spatial resolution, more limited field-of-view, and/or longer averaging time than the quantum interferometry technique presented here.  NV center spectroscopy leverages the unique capability of these optically-active defect spins to act as nanoscale sensors directly integrated with the diamond crystal matrix, allowing strain measurements that are simultaneously fast, high-resolution, and sensitive.

The NV center in diamond is a well-characterized  \cite{MansonNVstructure2006,AcostaTemperatureDependence2010,DohertyPhysicsReports2013} defect, currently used for many applications \cite{RondinMagnetometry2014,NVNanoReview2014,BalasubramanianNVBioimaging2014,NVMRIreview2019,MorevaEnsembleThermometry2020}.  It is an S=1 electronic system comprising a substitutional nitrogen impurity adjacent to a vacancy in the diamond carbon lattice.  Figure \ref{fig:introfigure}a gives a simplified diagram of the NV electronic and spin energy levels.  Exposure to green light excites the electronic state into a phonon sideband; it quickly decays via spontaneous emission of red fluorescence.  Due to a differential coupling to an alternative decay path, the fluorescence intensity depends on the NV spin state, and there is preferential decay to the $m_S=0$ sublevel.  In addition, the NV electronic spin can have long coherence lifetimes ($\mu$s to ms) under ambient conditions.  These key properties enable optical initialization and readout of the NV spin state, allowing fast, high-precision sensing of environmental parameters that shift the NV spin sublevels, including magnetic fields \cite{NVimaging2008,SingleNVMagSensing2008,NVMagnetometer2008}, strain \cite{StrainPaper,AusStrain2019,EnglundStrain2016}, and temperature \cite{LukinThermometry2013,AwschalomThermometry2013,WrachtrupThermometry2013}.  

The axis between the nitrogen and vacancy that make up an NV can be parallel to one of four crystallographic axes in the diamond crystal, creating four possible NV classes that are typically equally populated in an NV ensemble.  When a bias magnetic field $B_z$ is aligned with one of these NV classes, the NV ground state spin Hamiltonian for that class reduces to \cite{StrainPaper}:
\begin{equation}
\label{eqn:ham}
    H=(D+M_z)S_{z}^2+\gamma B_z S_z,
\end{equation}
where $D$ is the temperature-dependent zero-field splitting and $M_z$ is the axial stress-induced energy shift.  The detail of Fig.~\ref{fig:introfigure}a shows this ground-state energy level structure, together with the microwave (MW) frequencies used to probe NV spin transitions.  (Note that there is also hyperfine structure associated with these NV energy levels, which is not indicated in Fig. 1a.)

Previous diamond strain measurements using NV centers were performed via continuous wave (CW) \cite{StrainPaper,EnglundStrain2016,HoAnvilCells2020,HoAmorphousSolids2021} or pulsed \cite{AusStrain2019} optically detected magnetic resonance (ODMR) -- a simple, robust technique where laser and MW fields are applied to the diamond and the MW frequency is swept or modulated.  When the MW drive is resonant with the $\ket{0}\rightarrow\ket{\pm1}$ spin transitions, the output NV fluorescence decreases as population transfers between spin states.  (A more detailed discussion of CW-ODMR, including measurements with this technique of the diamond used in this work, is included in Appendix \ref{sec:CWODMR}.)

\section{Strain measurements using quantum interferometry}

We achieve a two order-of-magnitude improvement in diamond strain sensitivity over previous work by overcoming two limitations of CW-ODMR.  First, we perform a Ramsey interferometric measurement, applying temporally separated, pulsed MW fields to measure the phase acquired by the NV spins during a free evolution period.  This avoids spectral broadening induced by the MW fields, and fully leverages the quantum coherence of the NV center \cite{SensitivityReport,DegenQuantumSensing2017}.  (We note that pulsed ODMR also avoids this spectral broadening; however, it is in turn subject to other sensitivity-reducing effects, including Fourier broadening and inefficient hyperfine driving.  See \cite{SensitivityReport} for a detailed comparison of pulsed ODMR and Ramsey interferometric measurements.)

Second, we employ a variation on the CPMG (Carr-Purcell-Meiboom-Gill) spin control sequence that is optimized for NV-ensemble strain sensing: strain-CPMG, as shown in Fig.~\ref{fig:sequencefig}a.  The phase acquired during this sequence depends on $M_z$, but is leading-order insensitive to magnetic inhomogeneity due to impurity spins -- the leading source of dephasing for NV ensembles in low-strain diamond \cite{BauchDecoherence2020}.  The strain-CPMG sequence \cite{AwschalomThermometry2013,WangThermometry2015} \footnote{When used for single-NV thermometry in \cite{AwschalomThermometry2013,WangThermometry2015} this pulse sequence was referred to as the T-CPMG or thermal Carr-Purcell-Meiboom-Gill sequence.}, and other magnetic-field-insensitive sequences \cite{WrachtrupThermometry2013,KonzelmannCoopDRamsey2018}, have been used for single-NV thermometry, but have not previously been used with NV ensembles or to measure diamond strain.

A strain-sensitive, magnetic-field-insensitive measurement protocol is possible because the strain term in the Hamiltonian with an aligned bias magnetic field, equation \eqref{eqn:ham}, is quadratic in $S_z$, while the magnetic field term is linear.  The need for an axial bias field, to maintain this distinguishability, limits such a protocol to measuring $M_z$ for one NV class at a time, which is assumed to result from a linear combination of normal strain along the three principal axes of the NV's coordinate system \cite{BarfussHybridSystems2019}.  Such a single-NV-orientation measurement satisfies many applications that require only the position and magnitude of strains and strain features, including nanostructure fabrication optimization \cite{KnauerFabStrain2020}, strain probes in diamond cells and amorphous solids \cite{HoAnvilCells2020,HoAmorphousSolids2021}, failure analysis of diamond devices, and particle detection \cite{MarshallQST2021}.  In applications where additional information about the strain tensor is beneficial, the magnetic field could be realigned to measure $M_z$ for all four NV classes.  This would allow reconstruction of components of the strain tensor \cite{StrainPaper}, at the cost of a factor of 2 in sensitivity.

\sequencefig

In a strain-CPMG measurement, we address the NV spins with a pulsed MW field employing two tones near the two ground-state spin transition frequencies, as illustrated in Fig.~\ref{fig:introfigure}a.  Because the $S_z$ and $S_z^2$ Hamiltonian terms respectively yield differential and common-mode shifts in the $\ket{\pm1}$ spin energy levels, we express the MW drive frequencies in terms of these shifts together with differential and common-mode detunings: $\omega_\pm=D+M_z+\delta_{\rm cm}\pm(\gamma B+\delta_{\rm diff})$.  An initial $\pi$/2-pulse on the $\ket{0}\rightarrow\ket{-1}$ transition prepares the NV electronic spins in an equal superposition of those states, which accumulates phase at a rate proportional to $(D+M_z)-\gamma B_z$. Triplets of $\pi$-pulses swap the $\ket{-1}$ population into $\ket{+1}$, giving a superposition that accumulates phase proportional to $(D+M_z)+\gamma B_z$ (see Appendix \ref{sec:scpmg} for further discussion of the swap pulses). Over a full strain-CPMG sequence, with an even number of such swaps, the total B-dependent phase sums to zero and the final phase depends only on $D+M_z$. Similarly, the phase difference due to the differential detuning terms $\pm \delta_{\rm diff}$ also sums to zero, while the $\delta_{\rm cm}$ terms add constructively; a final $\pi$/2-pulse encodes the phase difference acquired from the common-mode detuning $\delta_{\rm cm}$ into the population of the $\ket{0}$ state, which can be read out via spin-dependent fluorescence.

The antisymmetric order of $\pi$-pulses within the swaps are chosen to cancel out phase acquired due to pulse imperfections and the nonzero $\pi$-pulse duration \cite{AwschalomThermometry2013}.  Alternating the phase of one MW tone in the final swap operation changes the axis of rotation of that swap, allowing us to choose $\pm$X or $\pm$Y as the overall phase of the sequence (in the rotating frame). 

NV ensemble fluorescence measured after a strain-CPMG sequence with free evolution time $\tau$ has the form
\begin{equation}
\label{eqn:fluor}
\begin{split}
    f_{\rm X}^\pm &= f_i(1\pm A e^{-\tau/T_D} \sin(2\pi \delta_{cm} \tau+\phi_0)),\\
    f_{\rm Y}^\pm &= f_i(1\pm A e^{-\tau/T_D} \cos(2\pi \delta_{cm} \tau+\phi_0)),
    \end{split}
\end{equation}
for a rotation around $\pm$X or $\pm$Y.  Here, $A$ is the measurement contrast, $T_D$ is the strain-CPMG dephasing time, and $f_i$ is the fluorescence from the unpolarized NV ensemble, all of which depend on diamond material properties and technical details of the apparatus \cite{SensitivityReport}.  $\delta_{cm}$ is the common-mode detuning for both MW tones, and $\phi_0$ is a constant phase offset acquired during the drive and swap MW pulses.  To normalize for fluctuations in laser power during a measurement, which affect $f_i$, we interlace strain-CPMG sequences with opposite overall phase (alternating between $+$X and $-$X), and extract the interferometric visibility $\nu$:
\begin{equation}
\label{eqn:suppvis}
    \nu=\frac{f_{\rm X}^+-f_{\rm X}^-}{f_{\rm X}^++f_{\rm X}^-}=A e^{-\tau/T_D} \sin(2\pi \delta_{cm} \tau+\phi_0).
\end{equation}
A measurement of $\nu$ versus phase evolution time $\tau$ is shown in Fig.~\ref{fig:sequencefig}b.  

To convert $\nu$ to a measurement of $M_z$, we measure a calibration curve such as that shown in Fig.~\ref{fig:sequencefig}c. We vary the common-mode MW detuning $\delta_{\rm cm}$ by sweeping both drive tones in the same direction (e.g., low to high frequency) while keeping a fixed evolution time $\tau_1$; this affects the fluorescence in the same way as a strain-induced change in $M_z$, giving an oscillation with period $1/\tau_1$ and amplitude $A e^{-\tau_1/T_D}$.  Fig.~\ref{fig:sequencefig}c also shows a calibration for the differential detuning $\delta_{\rm diff}$, where sweeping the two drive tones in opposite directions is equivalent to a change in the magnetic field strength.  Because $\delta_{\rm diff}$ and the magnetic field-induced precession cancel in the strain-CPMG sequence, this differential sweep has little effect on $\nu$.  

If the interference fringe amplitude $A e^{-\tau_1/T_D}$ is constant, then comparing the difference in visibility $\nu$ as measured at different spatial positions to a calibration curve gives a map of $M_z$.  (See below for a measurement protocol when this amplitude is not constant).  To convert this frequency measurement into strain, we follow \cite{BarfussHybridSystems2019} in expressing $M_z$ as a linear combination of normal strain:
\begin{equation}
    M_z=A_1 \epsilon_{zz}+A_2(\epsilon_{xx}+\epsilon_{yy}).
\end{equation}
We use the pure strain convention, common in previous NV-diamond strain measurements \cite{AusStrain2019,UdvarhelyiSpinStrain2018}, which give coupling constants $A_1=-8.0(5.7)$ GHz/strain and $A_2=-12.4(4.7)$ GHz/strain \cite{BarfussHybridSystems2019}.  For a quantitative evaluation of sensitivity to arbitrary strains, we take a weighted average and obtain $\bar{\epsilon}=-M_z$/(10.9(5.0) GHz/strain).  

As this procedure measures $D+M_z$ without distinguishing between the sample- and temperature-dependent $D$ and the strain shift $M_z$, it is best suited to measuring relative strain within a sample.  Diamond is an outstanding thermal conductor \cite{TwitchenThermalCond2001}, so $D$ can be considered uniform across the millimeter-scale sample \cite{WangThermometry2015}, while $M_z$ will be constant in time unless we apply an external stress.  We therefore interpret spatially differing measurements as revealing relative strain, whereas temporally differing measurements indicate changes in temperature.  One further limit is that a measurement at evolution time $\tau_1$ only determines the phase accumulated up to a factor of 2$\pi$; additional measurements at shorter evolution times can resolve this ambiguity, but for the material used in this study we find the strain variation to be within the dynamic range of a single $\tau$ value.

\section{Experimental demonstrations of interferometric strain microscopy}

\confocalfig

We first implemented the strain-CPMG protocol on diamond section A (see appendix \ref{sec:dephasingtimes}) using a confocal scanning laser microscope (illustrated in Fig.~\ref{fig:introfigure}b -- see also Appendix \ref{sec:confocalapparatus}), to quantitatively evaluate the achievable strain sensitivity. We evaluated the measurement precision as a function of averaging time via the Allan deviation \cite{AllanDeviation1966}, with results shown in Fig.~\ref{fig:confocalfig}b.  From this procedure, together with the measured confocal volume of 0.54(2) ${\rm \mu m^3}$, we obtain a volume-normalized strain sensitivity of $5(2)\times10^{-8} /\sqrt{\rm Hz \cdot \mu m^{-3}}$.  The measurement stability is limited by shot noise in the avalanche photodiode (APD) measurements of NV fluorescence; despite experimental imperfections, we obtain a sensitivity within 18\% of the calculated shot-noise limit (see Appendix \ref{sec:noiselimit}).  The dominant uncertainty in this measurement of strain sensitivity is the systematic uncertainty in the spin-strain coupling coefficients (discussed above).  Other sources of uncertainty, including statistical uncertainty in the Allan deviation measurement, uncertainty in the calibration curve fit used to convert visibility $\nu$ to frequency shift $M_z$, and uncertainty in the measured confocal volume, each contribute at the level of $\sim1\times10^{-9}.$

For measurement times greater than 20-30 seconds, determination of the visibility $\nu$ is limited by variation in the diamond's temperature of approximately 0.1 K/hour, due to drifts in lab temperature.  We therefore also measure using a ``gradiometry'' configuration: i.e., measurements alternate between the scan position and a reference position on the diamond surface (arbitrarily chosen for each scan), and the MW drive frequency is servoed on the reference measurement to compensate for thermal drift and preserve optimal sensitivity.  This protocol costs a factor of approximately $\sqrt{2}$ in sensitivity relative to a single-position protocol, due to the time spent making reference measurements.  However, it is stable for averaging times approaching $10^5$ seconds, allowing scanning strain measurements to achieve high precision without effort to control external temperatures beyond that in a typical lab environment.  The Allan deviation of measurements in this configuration is also shown in Fig.~\ref{fig:confocalfig}b. 

Figure \ref{fig:confocalfig} also shows two slices of a strain map produced by confocal scans across a feature in the diamond sample.  These measurements used the gradiometry configuration to counteract temperature drift during the scans, which required approximately 6 hours each.  Fig.~\ref{fig:confocalfig}a shows a map obtained by scanning across the diamond's surface, while panel Fig.~\ref{fig:confocalfig}c shows an orthogonal slice taken by scanning into the diamond.  As the strain magnitude grows and the feature's spatial extent shrinks with increasing depth of the confocal spot beneath the diamond surface, this feature may result from a scratch or extended crystallographic defect originating at or near the diamond's opposite surface.  

\imageronefig

When scanning at multiple depths below the diamond surface, as during the acquisition of Fig.~\ref{fig:confocalfig}c, the visibility $\nu$ cannot be converted to a measurement of $M_z$ by a single calibration curve, as the amplitude of the driving field from the planar MW loop falls off with depth, reducing the interference fringe amplitude $A e^{-\tau_1/T_D}$. We therefore measure what we term the ``XY-normalized'' visibility $\nu_{\rm{XY}}$:
\begin{equation}
\label{eqn:vxy}
    \nu_{\rm XY}=\frac{f_{\rm X}^+-f_{\rm X}^-}{\sqrt{(f_{\rm X}^+-f_{\rm X}^-)^2+(f_{\rm Y}^+-f_{\rm Y}^-)^2}}.
\end{equation}  
In addition to fluctuations in laser power, this quantity is also normalized against fluctuations in the interference fringe amplitude. 
Substituting equation \eqref{eqn:fluor} into equation \eqref{eqn:vxy} yields 
\begin{equation}
    \sqrt{(f_{\rm X}^+-f_{\rm X}^-)^2+(f_{\rm Y}^+-f_{\rm Y}^-)^2}=2f_{i}A e^{-\tau_1/T_D}
\end{equation}
as well as
\begin{equation}
\nu_{\rm XY}=\sin(2\pi \delta_{cm} \tau+\phi_0)).
\end{equation}
As a result, no amplitude calibration is required to extract $M_z$ from $\nu_{\rm{XY}}$.  Measuring X and Y quadratures costs a factor of $\sqrt{2}$ in sensitivity for a fixed overall measurement time; but the only cost from MW inhomogeneity is increased uncertainty for positions with smaller $A e^{-\tau_1/T_D}$, rather than erroneous systematic shifts in the measured strain.  The robustness of the XY-normalized technique is especially useful for widefield strain imaging (described below), where laser and MW power inhomogeneity can be significant \cite{KuDiracFlow2020}.

While scanned confocal microscopy allows strain mapping in three dimensions with a well-defined interrogation volume, many applications benefit from the speed and scalability of widefield imaging.  For example, diamond characterization and engineering require broad surveys of millimeter-scale diamond to understand the overall growth properties \cite{PurpleDiamond,Friel2009}; nanofabrication analysis could benefit from measuring many devices in parallel \cite{KnauerFabStrain2020}; and dark matter detection will require quickly analyzing large volumes of material to find particle impact sites \cite{MarshallQST2021}.  We therefore implemented wide-field NV-diamond strain imaging on diamond section B (see Appendix \ref{sec:dephasingtimes}) using a quantum diamond microscope (QDM), schematically illustrated in Fig.~\ref{fig:introfigure}c (see also Appendix \ref{sec:QDMapparatus}).  This QDM uses a Heliotis HeliCam lock-in camera, described in detail in \cite{DQ4R}, to quickly acquire NV fluorescence images.  Fig.~\ref{fig:imageronefig} shows widefield images of strain across a region of the diamond material.  Fig.~\ref{fig:imageronefig}a presents a broad survey comprising many individual fields-of-view, while Fig.~\ref{fig:imageronefig}b and c show single fields-of-view after one second of data acquisition.  Fig.~\ref{fig:imageronefig}b demonstrates the sensitive measurement of a typical low-strain region on the sample, while Fig.~\ref{fig:imageronefig}c shows the capability to image strain features of interest.  

In addition to preventing systematic shifts due to laser and MW power inhomogeneity, measuring $\nu_{\rm XY}$ resolves a second problem that is more pronounced with a QDM than a confocal microscope: strain gradients within a single pixel.  These gradients can be very large at the edges or centers of strain features, meaning measurement of the mean strain-induced frequency shift within a pixel may not adequately encode the strain structure.  Worse, gradients within a pixel yield a distribution of different precession frequencies, causing dephasing and reducing $T_D$ compared to low-gradient pixels -- potentially ``washing out'' the average strain shift and leading to noisy or mis-estimated strain readout.  However, the denominator of $\nu_{\rm XY}$ is proportional to $e^{-\tau_1/T_D}$, and therefore can be used as a signal identifying such high-gradient pixels.  Fig.~\ref{fig:imageronefig}d shows an example of this effect; the interference fringe amplitude $Ae^{-\tau_1/T_D}$ is reduced to approximately 1/$e$ of the value for nearby low-strain regions -- consistent with an additional strain gradient of about $1.4\times10^{-6}$ within the volume, and an attendant decrease in dephasing time from about 20 to 10 $\rm{\mu s}$.

Figure~\ref{fig:imageronefig}e shows a histogram of one-second Allan deviations for pixels in the low-strain QDM image, Fig.~\ref{fig:imageronefig}b.  The sensitivity variation is largely due to inhomogeneity of approximately 60\% in the green excitation laser power across the field-of-view -- lower laser power implies larger fractional shot noise, due to collection of fewer fluorescence photons.  Additionally, laser inhomogeneity can lead to reduced contrast $A$ in some regions, if the optical power is insufficient to fully repolarize the NV centers between strain-CPMG measurements.

Lacking depth restriction smaller than the focal plane of the 20X microscope objective used with the QDM, each pixel in the imaging field-of-view integrates light from a larger diamond volume than the confocal spot; Fig.~\ref{fig:imageronefig}e demonstrates the (non-volume-normalized) strain sensitivity achievable with this greater integration volume.  For diamond engineering applications requiring broad surveys of material (but not demanding the higher spatial resolution offered by the confocal method), the QDM's wide field-of-view and high sensitivity demonstrated in Fig.~\ref{fig:imageronefig} are significant advantages.  Additionally, the resilience of the strain-CPMG sequence to nuclear spin bath-induced magnetic inhomogeneity will allow sensitive strain characterization in diamonds with high nitrogen or $^{13}$C content, which currently is limited by broad ODMR linewidths and short dephasing times.  Future work towards other applications, such as photonic structure engineering and dark matter searches, will implement improved depth resolution in widefield imaging, e.g., using structured illumination or light-sheet microscopy \cite{MarshallQST2021}; and for each resulting ${\rm \mu m^3}$-scale pixel in a QDM image we expect the strain sensitivity to be similar to that measured with the confocal technique.  

\section{conclusion}

In conclusion, we demonstrated diamond crystal strain measurement with record volume-normalized sensitivity of $5(2)\times10^{-8}/\sqrt{\rm Hz \cdot \mu m^{-3}}$, using magnetic-field-insensitive quantum interferometry on an ensemble of NV centers.  Measurements with a scanning confocal microscope allow three-dimensional strain mapping with micron-scale spatial resolution, as well as quantitative characterization of the volume-normalized sensitivity.  Such a scanning protocol could enable sensitive, in situ measurements of strain in diamonds embedded in transparent materials or anvil cells.  The protocol also exceeds the sensitivity benchmark of $10^{-7}/\sqrt{\rm{Hz \cdot \mu m^{-3}}}$ for detection of particle-induced crystal damage in a directional search for WIMP dark matter \cite{MarshallQST2021}.  Measurements with a widefield imaging quantum diamond microscope (QDM) demonstrate speed and field-of-view that also surpasses a benchmark of (125$\times$125) $\rm{\mu m^2}$/sec for the proposed future dark matter detector \cite{MarshallQST2021}.  Beyond this basic physics application, widefield diamond strain imaging as demonstrated here will enable sensitive surveys of strain features for diamond engineering, as well as improved characterization of diamond nanofabrication for quantum and other devices.  In particular, short NV electronic spin dephasing times currently inhibit sensitive strain characterization in diamonds with high nitrogen or $^{13}$C content, which are used for many scientific and industrial applications including diamond knives and anvil cells.  The quantum interferometry technique is less sensitive to such sources of NV spin dephasing, and thus could help improve strain characterization and production or characterization of these and other diamond devices beyond high-purity CVD samples.

We note that the sensitivity of the confocal scanning measurements could be further enhanced with improvements to low-light detection by replacing the avalanche photodiode with (for example) a multi-pixel photon counter array; while future work will improve widefield imaging by increasing depth resolution.  Additionally, the strain sensitivity in both the confocal and QDM configurations could be further enhanced with improvements in diamond material, if the NV density can be increased without compromising a long strain-CPMG dephasing time $T_D$.

\section{Acknowledgements}
This work was supported by the DOE QuANTISED program under Award No. DE-SC0019396; the Army Research Laboratory MAQP program under Contract No. W911NF-19-2-0181; the DARPA DRINQS program under Grant No. D18AC00033; the DOE fusion program under Award No. DE-SC0021654; and the University of Maryland Quantum Technology Center.  This work was performed in part at the Harvard Center for Nanoscale Systems (CNS), a member of the National Nanotechnology Coordinated Infrastructure Network (NNCI), which is  supported by  the  National  Science  Foundation  under NSF award no. 1541959.

\appendix

\section{Interferometric strain sensing protocol}
\label{sec:scpmg}
%The TCPMG sensing protocol, similar to other Ramsey-like interferometric quantum sensing protocols, can be divided into three parts:
%\begin{itemize}
%    \item State preparation and readout: The NV spin state is read out and reset to $m_s=0$ state by illuminating with green light and collecting the resultant red fluorescence.
%    \item Coherent state manipulation: Microwave drives address the NV spin transitions to prepare or alter the state populations.
%    \item Free evolution intervals: In the absence of control fields, , environment-dependent phase.
%\end{itemize}
A schematic of the NV-diamond strain Carr-Purcell-Meiboom-Gill (strain-CPMG) sensing protocol is shown in Fig.~\ref{fig:SMfig1}a. First, green laser illumination prepares the NV ensemble ground-state spins in the $\ket{{\rm m_s} = 0}$ state. Then, a microwave (MW) $\pi/2$ pulse resonant with the $\ket{0}\rightarrow\ket{-1}$ transition prepares the NV ensemble in an equal superposition of the $\ket{0}$ and $\ket{-1}$ states. During the subsequent free evolution time (illustrated by red traces in the lower panel of Fig. \ref{fig:SMfig1}a), the NV spins accumulate a relative phase with the rate $D + M_z - \gamma B_z$, where these parameters are defined in the main text in the context of the NV spin Hamiltonian. Next, triplets of MW $\pi$-pulses collectively swap the NVs between the $\ket{-1}$ and $\ket{+1}$ states (see below for more details). Thus, during the subsequent free evolution time the phase accumulation rate is $D + M_z + \gamma B_z$. Dividing the total free evolution time evenly between $\ket{+1}$ and $\ket{-1}$ states yields a final accumulated NV phase independent of $B_z$ in a form of balanced spin-state interferometry designed to be exclusively sensitive to $D+M_z$. The final $\pi/2$ pulse converts the accumulated phase into fractional NV population in $\ket{0}$ and $\ket{-1}$, which is read out via state-dependent fluorescence.

\smfive

\subsection{Time evolution during swap pulse}
During the strain-CPMG sequence, triplets of microwave (MW) $\pi$-pulses collectively swap the NV population between the $\ket{-1}$ and $\ket{+1}$ states (see Fig. \ref{fig:SMfig1}a). In this section, we calculate the time evolution of an arbitrary initial state during the swap operation.  To demonstrate the dynamics of the swap operation, we derive an analytic solution for the ideal case of resonant MW drive fields; the more general case of driving with detunings must be treated numerically \cite{MaminNanoNMR2013}. The NV spin Hamiltonian under resonant MW driving is

\begin{equation}
\begin{split}
    H = (D+M_z) S_z^2 + \gamma B S_z  &+ \\  [\gamma B_+ \cos(\omega_+ t + \Delta \phi_+) &+ \gamma B_- \cos(\omega_- t + \Delta \phi_-)]S_x\,,
    \end{split}
\end{equation}

where $\omega_\pm \equiv D+M_z\pm\gamma B$, $B_\pm$, and $\Delta \phi_\pm$ are the frequencies, amplitudes, and phases of the applied MW fields. 
After performing a unitary transformation $V = {\rm diag}(e^{-i \omega_+},1,e^{-i \omega_-})$ into a frame rotating with the MW drive fields, and dropping rapidly oscillating terms, this system can be solved analytically. The evolution operator in the transformed frame is then:
\begin{widetext}
\begin{equation}
\label{eqn:evulution}
U(B_\pm,\Delta \phi_\pm;t) = \left(\begin{matrix}
\frac{B_-^2 + B_+^2 cos(\omega_e t)}{B_+^2 + B_-^2} & \frac{-i B_+ e^{-i \Delta \phi_1}sin(\omega_e t)}{\sqrt{B_+^2 + B_-^2}} & \frac{(-1+cos(\omega_e t)) B_+ B_- e^{-i (\Delta \phi_1-\Delta \phi_2)} }{B_+^2 + B_-^2}\\
\frac{-i B_+ e^{-i \Delta \phi_1}sin(\omega_e t)}{\sqrt{B_+^2 + B_-^2}} & cos(\omega_2 t) & \frac{-i B_- e^{-i \Delta \phi_2}sin(\omega_e t)}{\sqrt{B_+^2 + B_-^2}} \\ 
\frac{(-1+cos(\omega_e t)) B_+ B_- e^{-i (\Delta \phi_1-\Delta \phi_2)}}{B_+^2 + B_-^2} & \frac{-i B_- e^{-i \Delta \phi_2}sin(\omega_e t)}{\sqrt{B_+^2 + B_-^2}} & \frac{B_+^2 + B_-^2 cos(\omega_e t)}{B_+^2 + B_-^2} 
\end{matrix}\right)\,,
\end{equation}
\end{widetext}
where $\omega_e\equiv\frac{\gamma}{2\sqrt{2}}\sqrt{B_+^2 + B_-^2}$ \cite{MaminNanoNMR2013}. After solving the evolution of an initial state $\tilde{\ket{\psi(0)}}$ in the transformed frame, we transform the state back to the lab frame: $\ket{\psi}_{\rm lab} = V \tilde{\ket{\psi}}$. For a detailed derivation of the evolution operator refer to Ref. \cite{MaminNanoNMR2013}; for an alternative discussion of strain-CPMG swap pulses, see Ref. \cite{AwschalomThermometry2013}. 

We construct the state evolution under the first swap operation in Fig. \ref{fig:SMfig1}a (i.e., swap from $\ket{-1}$ to $\ket{+1}$) using the evolution operator:

\begin{align}
\label{eqn:swap_evolution}
    U_{\rm swap}(t)=
        \begin{cases}
        U(B_-=1,B_+=0,\Delta \phi_\pm=0;0<\rm{t<t_\pi})\\
        U(B_-=0,B_+=1,\Delta \phi_\pm=0;\rm{t_\pi}<t<\rm{2t_\pi})\\
        U(B_-=1,B_+=0,\Delta \phi_\pm=0;\rm{2t_\pi}<t<\rm{3t_\pi})
        \end{cases}
\end{align}

Fig. \ref{fig:SMfig1}b shows the population in each state as a function of time during these swap pulses, for an ideal strain-CPMG measurement, while Fig. \ref{fig:SMfig1}c shows the time evolution for an arbitrary initial state under $U_{\rm swap}(t)$, to better illustrate the full swap dynamics with unequal populations.

\subsection{Strain-CPMG with N$>2$ swap pulses}

We measure the strain-CPMG dephasing time $T_D$ using strain-CPMG sequences with increasing numbers of swap operations.  While strain-CPMG sequences cancel phase accumulated due to magnetic fields that are static or slowly varying over the timescale of the sequence, these sequences remain vulnerable to higher frequency magnetic noise (e.g., dephasing induced by electronic or nuclear spin bath dynamics). In the absence of all other contributions, $\rm{T_D}$ would therefore be expected to approach the ensemble decoherence time ($\rm{T_2}$), and increase with additional swap operations.  However, as shown in Fig. \ref{fig:SMfig2}a-c, $\rm{T_D}$ is largely independent of the number of swap-operations applied, and shorter than even the Hahn echo $\rm{T_2}$ $\approx100\rm{\mu s}$ observed in this diamond.  This suggests that $\rm{T_D}$ for the 2-swap sequence (and sequences with increasing number of swap operations) is dominated by inhomogeneous common-mode shifts, effects that the strain-CPMG sequence is designed to preserve -- for example, inhomogeneous strain distributions within a confocal spot or camera pixel, potentially arising from microscopic variations in the distribution of nitrogen or other impurities near each NV center.  Meanwhile, as demonstrated by the decreasing interference fringe amplitude $A$ in Figs.~\ref{fig:SMfig2}a-c, increasing numbers of swap operations allow small MW pulse errors to accumulate, reducing measurement contrast; measurements in this work were therefore performed using a strain-CPMG sequence with two swap operations.

\smsix

\section{Comparison of dephasing times}
\label{sec:dephasingtimes}

The single-crystal bulk diamond material used in this work was grown by Element Six, Inc, via \{100\} chemical vapor deposition (CVD) employing isotopically purified $^{12}$C enriched with nitrogen. After growth, diamond sections A and B were cut from the material and separately electron-irradiated and annealed, forming NV centers with estimated [NV] density of 0.3 parts per million in section A (used for confocal measurements) and 0.5 parts per million in section B (used for widefield imaging).  The resulting octagonal diamond chips have dimensions of $3.4\times3.4\times0.1$ mm$^3$ and feature relatively homogeneous crystal strain (see Fig.~\ref{fig:imageronefig} of the main text).  Both diamond sections have a similar ensemble NV inhomogeneous dephasing time $T_2^\star\approx7.5\mu$s and strain-CPMG dephasing time $T_D\approx20\mu$s.

In addition to the strain-CPMG dephasing time $T_D$ (discussed in the main text and shown in Fig.~\ref{fig:SMfig3}a), we also measure the inhomogeneous dephasing time $T_2^*$ using a standard, single quantum (SQ) Ramsey measurement protocol \cite{SensitivityReport}. This protocol treats the NV ground state spin as an effective spin-1/2 system, only addressing one of the two transitions $\ket{0} \leftrightarrow \ket{+1}$ or $\ket{0} \leftrightarrow \ket{-1}$ (which are split by the applied magnetic bias field). Fig. \ref{fig:SMfig3} shows measured visibility free induction decays for the strain-CPMG and Ramsey protocols. The three frequencies visible in the Ramsey curve correspond to drive detunings from each of the three hyperfine levels; these three frequencies are not present in the strain-CPMG measurement because the hyperfine splitting arises from an effective magnetic field and cancels under that protocol. This comparison clearly demonstrates the enhanced dephasing time attained with strain-CPMG compared to a standard Ramsey measurement.

\smseven

\section{Confocal microscope}
\label{sec:confocalapparatus}

For the confocal measurements reported in this work, we used a custom-built scanning confocal laser microscope optimized for NV spin-state manipulation and optical measurements.  A diode-pumped solid-state laser provides 10 mW of 532 nm excitation light at maximum power, which is switched by an acousto-optic modulator (AOM) and spatially filtered by a polarization-maintaining single-mode fiber.  (The SNR for NV spin-state determination continued to improve up to the maximum available power.)   This light is reflected by a dichroic mirror into the aperture of a microscope objective (Olympus, 100X magnification, 0.9 NA air), and focused into the diamond sample, which is held on a nanopositioning stage.  Red fluorescence from the NVs is collected by the same objective, and passes through the dichroic mirror into the collection optics.  A pinhole spatially filters the fluorescence light, which is then focused onto an avalanche photodiode (APD) operating in linear mode.  Microwave (MW) control of the NV spins is provided by two MW signal generators, each coupled through a 16 watt amplifier joined by a power combiner; the combined MW signal is broadcast to the NVs by an omega-loop antenna fabricated on a glass coverslip held in contact with the diamond sample.  MW- and laser-induced temperature change within each measurement cycle was negligible, but we did observe changes in the sample temperature on the order of tens of mK as a function of duty cycle; we controlled this effect by keeping the confocal duty cycle constant throughout all of the measurements reported here.

To quantitatively characterize the NV ensemble within the diamond volume addressed by the confocal microscope and determine the NV ensemble's volume-normalized strain sensitivity, we measured the confocal point spread function using a single NV center, approximately 1 $\rm{\mu m}$ below the surface of a different diamond sample of similar geometry.  For this measurement, we placed a single photon counting module (SPCM) at the focus of the collection optics, switching between SPCM and APD by means of a flip mirror.  Fitting the measured fluorescence distribution to a three-dimensional Gaussian function and integrating over its volume yielded the confocal volume of 0.54(2) $\rm{\mu m^3}$.

\section{QDM widefield imager using heliCam C3 lock-in camera}
\label{sec:QDMapparatus}

For the widefield imaging reported in this work, we used a quantum diamond microscope (QDM) apparatus and Heliotis heliCam C3 camera -- both described in detail in Ref.~\cite{DQ4R}.  The heliCam C3 subtracts alternating NV fluorescence exposures in hardware, and digitizes only their difference, allowing the measurement contrast to fill each pixel's 10-bit dynamic range. The rate of these exposures is defined by a user-controlled demodulation signal at frequency $f_{\rm{demod}}$; during one demodulation cycle, the camera takes four exposures and digitizes two difference images.  $N_{\rm demod}$ of these cycles are accumulated and their average is sent to the computer.  Typical values for this work ($N_{\rm demod} = 24$ and $f_{\rm demod} = 6.5 $ kHz) give an external frame rate of approximately 270 Hz.

The heliCam adds an independent, fixed pixel-by-pixel offset to each exposure in a demodulation cycle, likely arising from the different capacitor banks used to store the exposures for subtraction. In Ref. \cite{DQ4R}, this offset was subtracted with a separate calibration measurement for each field-of-view on the diamond, in preparation to sense magnetic fields due to an additional sample signal.  The strain-sensing applications addressed here typically require quickly imaging many diamond fields-of-view, and preclude a full calibration measurement at each one.  We therefore subtracted out these offsets by taking the difference between two frames at drive frequencies on either side of a Ramsey fringe; we measured for 0.5 seconds at each frequency to give an overall one-second data acquisition time.

In contrast to the confocal measurements, we found no evidence of MW- or laser-induced heating in the QDM configuration.  This is likely due to reduced thermal contact between the MW loop and diamond sample in the QDM, where the MW loop is freestanding, compared to the confocal, where the MW loop is printed on a coverslip mounted directly against the sample surface.

We note that while the greater interrogation volume in the wide-field imaging setup demonstrated here improves measurement times for applications not requiring sub-micron spatial resolution, it necessarily implies reduced rejection of out-of-focus light compared to the confocal geometry.  Widefield diamond imaging thus brings with it the possibility of measurement contamination by collection of light that has internally reflected from elsewhere in the diamond.  We see no evidence of this effect either in the measurements presented here or previous widefield QDM measurements able to reliably resolve micron-scale magnetic or strain features  \cite{DQ4R, QDMreview,GlennGGG,StrainPaper,GlennMTB2015,OpticalMagneticImagingLivingCells}.  In the current work, the imaging of high- and low-strain regions within microns of each other (as shown in Fig.~\ref{fig:imageronefig} of the main text) indicate that optical measurement contamination, if present, is not significant enough to blur the sharp edges of strain features in the diamond sample.

\section{Noise limit on strain sensitivity}
\label{sec:noiselimit}
To determine the expected noise limit to sensitivity in the confocal measurements, we estimate the noise current and voltage from the avalanche photodiode (APD) and Johnson noise in the transimpedance amplifier; convert this estimated noise from voltage to visibility $\nu$; and finally, use a strain calibration curve to determine the shot noise limit on our strain measurements.

For fluorescence detection in the confocal microscope, we used a Thorlabs APT410A APD-based photodetector.  The RMS noise current in an APD is given by
\begin{equation}
\label{eqn:noise_current}
    i_{\rm N} = \sqrt{2q\left[I_{\rm DS} + \left(I_{\rm DB} M^2 + R_0(\lambda)M^2 P_{\rm s}\right) F\right]B}\,, 
\end{equation}
where the first term is surface dark current; the second term is bulk leakage current; and the third term is photon shot noise from NV fluorescence.  Both the bulk leakage current and the photon shot noise are enhanced by the APD gain $M$ and excess noise factor $F$.  Additionally, $q$ is the electron charge; $R_0(\lambda)$ is the APD responsivity at $M=1$ for a given wavelength; $P_{\rm s}$ is the incident optical power onto the APD; and $B$ is the measurement bandwidth.

The responsivity at $M=100$ for 700 nm is approximately 45 A/W; combined with a transimpedance gain of 250 kV/A, the total gain is $11.25 \times 10^6$ V/W. A typical APD ouptut voltage of 5.2 mV thus corresponds to fluorescence power $P_{\rm s}$ = 0.46 nW, giving $R_0(\lambda)M^2 P_{\rm{s}} \approx 0.21$ mA. 

The typical noise equivalent power (NEP) for the APD410A with no input light is specified as 0.04 pW/$\sqrt{\rm{Hz}}$; cases where the entire NEP is due to surface or bulk dark current give upper limits of $I_{\rm{DS}} \leq 200$ pA and $I_{\rm{DB}}\leq 2$ pA (obtained by multiplying NEP by responsivity and gain if appropriate). These are each negligible compared to the photon shot noise term, even after accounting for the factor of $M^2$ in the bulk dark current term.

\smeight

The excess noise factor for the APD410A is $F=4$. Finally, we measured the DAQ system bandwidth $B$ to be 700 kHz, via the rise time of the APD response to a fast AOM pulse.  (The AOM rise time was measured separately, with a high-power photodiode and fast oscilloscope, and shown to be significantly faster than the DAQ system bandwidth.)  Eq. \eqref{eqn:noise_current} thus gives an expected RMS noise current of $i_{\rm N} = 1.4 ~{\rm nA}$. The transimpedance gain of 250 kV/A then yields the estimated RMS noise voltage:
\begin{equation}
\label{eqn:SN_noise}
    v_{\rm SN} = 0.34 ~{\rm mV}\,.
\end{equation}

The estimated RMS Johnson-Nyquist noise in the transimpedance amplifier is given by $v_{\rm JN} = \sqrt{4 k_{\rm B} T R B}$, where $k_{\rm B}$ is the Boltzmann constant, $T$ is the temperature ($\approx 300$ K), $R$ is the resistor value, and $B$ is the bandwidth. The equivalent resistance of the APD410A's transimpedance amplifier is 250 ${\rm k\Omega}$, resulting in an estimated RMS Johnson-Nyquist noise of
\begin{equation}
    v_{\rm JN} = 0.05 ~{\rm mV}\,.
\end{equation}
$v_{\rm{JN}}$ adds in quadrature with $v_{\rm{SN}}$, and therefore is essentially negligible; the expected RMS voltage noise is thus given by Eq. \eqref{eqn:SN_noise}.  Digitization noise and error, respectively specified at 15 $\rm{\mu}$V RMS and 52 $\rm{\mu}$V absolute accuracy for the NI PCIe-6251 DAQ used in this work, are also negligible compared to shot noise.

The uncertainty in the visibility $\nu=\frac{f_{\rm X}^+-f_{\rm X}^-}{f_{\rm X}^++f_{\rm X}^-}$ is given by
\begin{equation}
    \sigma(\nu) = \sqrt{\left(\frac{1 -\nu}{ f_{\rm X}^++f_{\rm X}^-} \right)^2 \, \sigma(f_{\rm X}^+)^2 + \left(\frac{1 + \nu}{f_{\rm X}^++f_{\rm X}^-} \right)^2 \, \sigma(f_{\rm X}^-)^2}
\end{equation}
Noting that $\nu \propto A e^{-\tau_1/T_D} \sim \mathcal{O}(0.01) \ll 1$, we have
\begin{equation}
    \sigma(\nu) \simeq \frac{\sqrt{\sigma(f_{\rm X}^+)^2 + \sigma(f_{\rm X}^-)^2}}{f_{\rm X}^+ + f_{\rm X}^-} = \frac{\sqrt{2} \, v_{\rm SN}}{f_{\rm X}^+ + f_{\rm X}^-} \approx 0.046\,.
\end{equation}
Converting this signal uncertainty to strain units using a fitted calibration curve (see Fig.~\ref{fig:sequencefig}b and associated discussion in the main text) gives an estimated noise limit on strain sensitivity of $5.2 \times 10^{-8}/\sqrt{\rm{Hz}}$ (the blue dotted line in Fig.~3b of the main text).  This noise limit is approximately 18\% smaller than the measured Allan deviation at one second shown in Fig.~\ref{fig:sequencefig}b, $6.4(1) \times 10^{-8}$.  The shot noise calculation presented here is not normalized to the confocal spot size; assuming a square-root scaling with volume, the shot noise limit and measured sensitivity are, respectively, 3.8 and 4.7(1) $\times10^{-8}/\sqrt{\rm{Hz}\cdot\rm{\mu m^{-3}}}$.  Note that the uncertainty in the spin-strain coupling coefficient is common to both the shot noise calculation and measured sensitivity, and does not affect the comparison between them.

\section{CW-ODMR strain imaging}
\label{sec:CWODMR}

CW-ODMR is a robust measurement method where continuous wave laser and MW fields are applied to the diamond.  When the MWs are off-resonance, the green laser light polarizes the NV center electron spins into the $\ket{0}$ state, maximizing their fluorescence output.  When the MWs are resonant with the ground-state spin transitions, they mix the $\ket{0}$ and $\ket{+1}$ or $\ket{-1}$ states, reducing the fluorescence.  Sweeping the MW frequencies and monitoring the fluorescence allows determination of $\ket{0}\rightarrow\ket{\pm1}$ transition frequencies.

The NV ground state spin Hamiltonian with a bias magnetic field $|\vec{B}|>1$ mT, as used in this work, reduces to \cite{StrainPaper}
\begin{equation}
\label{eqn:vectorhamiltonian}
    H_\kappa = (D + M_{z,\kappa})S_{z,\kappa}^2+\gamma \vec{B}\boldsymbol{\cdot}\vec{S}_\kappa,
\end{equation}
where $\kappa=(1,2,3,4)$ represent the four possible classes of NV centers pointing along four crystallographic directions (as shown in Fig.~\ref{fig:SMfig_odmr}a).

As discussed in the main text, in this work we align the bias magnetic field with one of the four NV classes, which enables separation of strain $(\propto S_z^2)$ and magnetic field $(\propto S_z)$ terms in the Hamiltonian for that NV class.  The three remaining NV classes each therefore see equal magnetic field projections $\vec{B}\boldsymbol{\cdot}\vec{S}$.  Figure \ref{fig:SMfig_odmr}c shows an example CW-ODMR spectrum under this condition, acquired during this work.  The aligned NV class sees an axial field $B_z\approx2.1$ mT, while the projection onto each other class is $\vec{B}\boldsymbol{\cdot}\vec{S}\approx0.75$ mT.  Each NV class yields two dips in fluorescence in the spectrum, corresponding to the $\ket{0}\rightarrow\ket{\pm1}$ spin transitions; each of these transitions is split into three lines by the hyperfine interaction with the spin-1 $^{14}$N nucleus.  

The transition frequencies for the aligned NV class are 
\begin{equation}
\label{eqn:odmrfreqs}
    f_{\pm}=(D+M_z)\pm\gamma B_z,
\end{equation}
where $\gamma$ is the NV gyromagnetic ratio.  Measuring both the $\ket{0}\rightarrow\ket{-1}$ and $\ket{0}\rightarrow\ket{+1}$ transitions allows separation of frequency shifts induced by magnetic field and strain or temperature; magnetic fields induce a differential splitting between the two resonances, while strain shifts them both in common mode (as discussed in the main text).  For CW-ODMR strain imaging, a camera acquires a widefield image of fluorescence intensity at each applied MW drive frequency.  Combining these images gives a MW spectrum for each pixel in the camera field-of-view; these spectra are fitted to Lorentzian lineshapes to determine the frequencies of equation \eqref{eqn:odmrfreqs} and obtain maps of $(D+M_z)$ and $B_z$.  The temperature-dependent zero-field splitting $D$ and the axial stress-induced frequency shift $M_z$ are then separated as in the main text, by assuming the temperature to be constant across the field-of-view and the strain to be constant with time; subtracting the mean value across a field-of-view thus gives a map of relative strain in the diamond.  Fig.~\ref{fig:SMfig_odmr}b shows one such CW-ODMR map, of diamond section B used in this work.  Pixels with large strain gradients, like the region shown in Fig.~\ref{fig:imageronefig}d of the main text, also affect CW-ODMR, in this case via deviations from Lorentzian lineshapes which interfere with the fitting procedure; Fig.~\ref{fig:SMfig_odmr}d shows a map of fit residuals for the strain map of Fig.~\ref{fig:SMfig_odmr}c.  

The measurement parameters used and strain sensitivity achieved in this CW-ODMR measurement both parallel those previously published in \cite{StrainPaper}.  In those previous measurements, a strain measurement precision of approximately $5\times10^{-8}$ was reached with a three-hour averaging time using a 13-$\mu$m-thick NV-rich diamond overgrowth layer.  The optimal lateral spatial resolution reported in those measurements was 1 $\mu$m, giving a minimal interrogation volume of 13 $\mu$m$^3$ per pixel and a corresponding volume-normalized strain sensitivity of $2\times10^{-5}/\sqrt{\rm{Hz}\cdot\rm{\mu m}^{-3}}$.

For further discussion of CW-ODMR strain imaging, including simultaneous vector measurements of all four NV classes, see \cite{StrainPaper}.

%\reftitle{References}
%\bibliographystyle{apsrev4-1}
\bibliography{MasonNVbib.bib}

\clearpage

\end{document}